\documentclass[prb,twocolumn,showpacs,superscriptaddress]{revtex4-1}
\usepackage{amsmath}
\usepackage{amssymb}
\usepackage{graphicx}
\usepackage{dcolumn}
\usepackage{bm}

\begin{document}

\title{Magnetic and defect probes of the SmB$_6$ surface state}

\author{Lin Jiao}\email{Lin.Jiao@cpfs.mpg.de}
\affiliation{Max-Planck-Institute for Chemical Physics of Solids,
N\"othnitzer Str. 40, 01187 Dresden, Germany}
\author{Sahana R\"o{\ss}ler}
\affiliation{Max-Planck-Institute for Chemical Physics of Solids,
N\"othnitzer Str. 40, 01187 Dresden, Germany}
\author{Deepa Kasinathan}
\affiliation{Max-Planck-Institute for Chemical Physics of Solids,
N\"othnitzer Str. 40, 01187 Dresden, Germany}
\author{Priscila F. S. Rosa}
\affiliation{Department of Physics and Astronomy, University of
California, Irvine, CA 92697, USA}
\affiliation{Los Alamos National Laboratory, Los Alamos, NM87545, USA}
\author{Chunyu Guo}
\affiliation{Center for Correlated Matter and Department of Physics,
Zhejiang University, Hangzhou 310058, People's Republic of China}
\author{Huiqiu Yuan}
\affiliation{Center for Correlated Matter and Department of Physics,
Zhejiang University, Hangzhou 310058, People's Republic of China}
\author{Chao-Xing Liu}
\affiliation{Dept.\ of Physics, The Pennsylvania State University,
University Park, Pennsylvania 16802, USA}
\author{Zachary Fisk}
\affiliation{Department of Physics and Astronomy, University of
California, Irvine, CA 92697, USA}
\author{Frank Steglich}
\affiliation{Max-Planck-Institute for Chemical Physics of Solids,
N\"othnitzer Str. 40, 01187 Dresden, Germany}
\affiliation{Center for Correlated Matter and Department of Physics,
Zhejiang University, Hangzhou 310058, People's Republic of China}
\author{S. Wirth}\email{wirth@cpfs.mpg.de}
\affiliation{Max-Planck-Institute for Chemical Physics of Solids,
N\"othnitzer Str. 40, 01187 Dresden, Germany}

\date{\today}

\begin{abstract}
The impact of non-magnetic and magnetic impurities on topological
insulators is a central problem concerning their fundamental physics and
possible novel spintronics and quantum computing applications. SmB$_6$,
predicted to be a topological Kondo insulator, is considered a benchmark
material. Using a spin-polarized tip in scanning tunneling spectroscopy
destroys the signature peak of the topological surface state, revealing
its spin texture. Further, combining local STS with macroscopic transport
measurements on SmB$_6$ containing different substitutions enables us to
investigate the effect of impurities. The surface states around impurities
are locally suppressed with different length scales depending on their
magnetic properties and, for sufficiently high impurity level, globally
destroyed. Our study points directly to the topological nature of
SmB$_6$, and unveils, microscopically and macroscopically, how impurities
-- magnetic or non-magnetic -- affect topological surface states.
\end{abstract}
\maketitle

Topological surface states (TSS) are novel quantum electronic states which
not only serve as a playground for realizing many exotic physical phenomena
(such as magnetic monopoles, Majorana fermions, and the quantum anomalous
Hall effect) but may also find several fascinating applications, e.g. in
spintronics or quantum computing \cite{fuk08}. Remarkably, these states are
theoretically predicted to be robust against backscattering from non-magnetic
impurities due to their chiral spin texture, whereas a magnetic impurity
breaks time-reversal symmetry and therefore induces spin scattering. Within
the framework of the Anderson impurity model, however, the spin of a magnetic
impurity may also be screened by the conduction electrons via the Kondo
effect, resulting in an effectively non-magnetic scattering center at
sufficiently low temperature \cite{zit10}. Conversely, nonmagnetic impurities
in a Kondo lattice may generate magnetic scattering by locally increasing the
density of states or by creating a ``Kondo hole'' in the lattice. This calls
for a detailed understanding of the impact of impurities -- non-magnetic and
magnetic -- on TSS.

SmB$_6$ was theoretically predicted to be a topological Kondo insulator in
which a direct bulk gap is induced by Kondo hybridization, and the TSS reside
in this small bulk gap \cite{dze10,tak11,lu13,ale13}. Compared to weakly
correlated topological insulators, SmB$_6$ features a small, but fully opened
bulk gap \cite{eo18}. Therefore, the TSS dominate the density of states (DOS)
at Fermi level $E_{\rm F}$ at low temperature, such that insulating bulk and
metallic surface properties can be well distinguished in spectroscopic
\cite{jiao16} and transport measurements \cite{kim14}. In addition, the small
bulk gap can easily be suppressed by doping, converting the system to a
trivial insulator \cite{schu12}. These aspects provide an excellent starting
point to study the effects of impurities in SmB$_6$.

Although the metallic nature of the surface states in SmB$_6$ has been
confirmed by several experiments \cite{wol13,kim13,kim14,chen15}, pinning
down their topological origin remains challenging and controversial
\cite{miy12,zhu13,hla15}. Consequently, detecting the spin-texture of the
metallic surface states is cardinal and crucial. Considerable efforts have
been made to uncover the helical spin texture of the surface state by
spin-resolved ARPES \cite{nxu14,sug14} and spin injection \cite{song16}.
However, the surface conditions are multifarious \cite{roe16}. Therefore,
investigations on well defined and non-reconstructed surfaces are important
and a microscopic probe is called for. In this respect, scanning tunneling
microscopy and spectroscopy (STM/S) studies have clearly demonstrated their
ability to characterize the bulk and surface state band of SmB$_6$
\cite{yee13,roe14,ruan14}. In particular, a major contribution to the
dominating peak in the low-$T$ tunneling spectra at a bias voltage
$V_{\rm b} \approx -6.5$ mV was related to the surface state \cite{jiao16}.
Here we employ spin-polarized STS to illustrate the existence of a surface
spin texture. The local and global impact of non-magnetic and magnetic
impurities on the surface states is further studied by comparing
microscopic STS with macroscopic transport measurements.

\begin{figure}
\centering
\includegraphics[width=8cm]{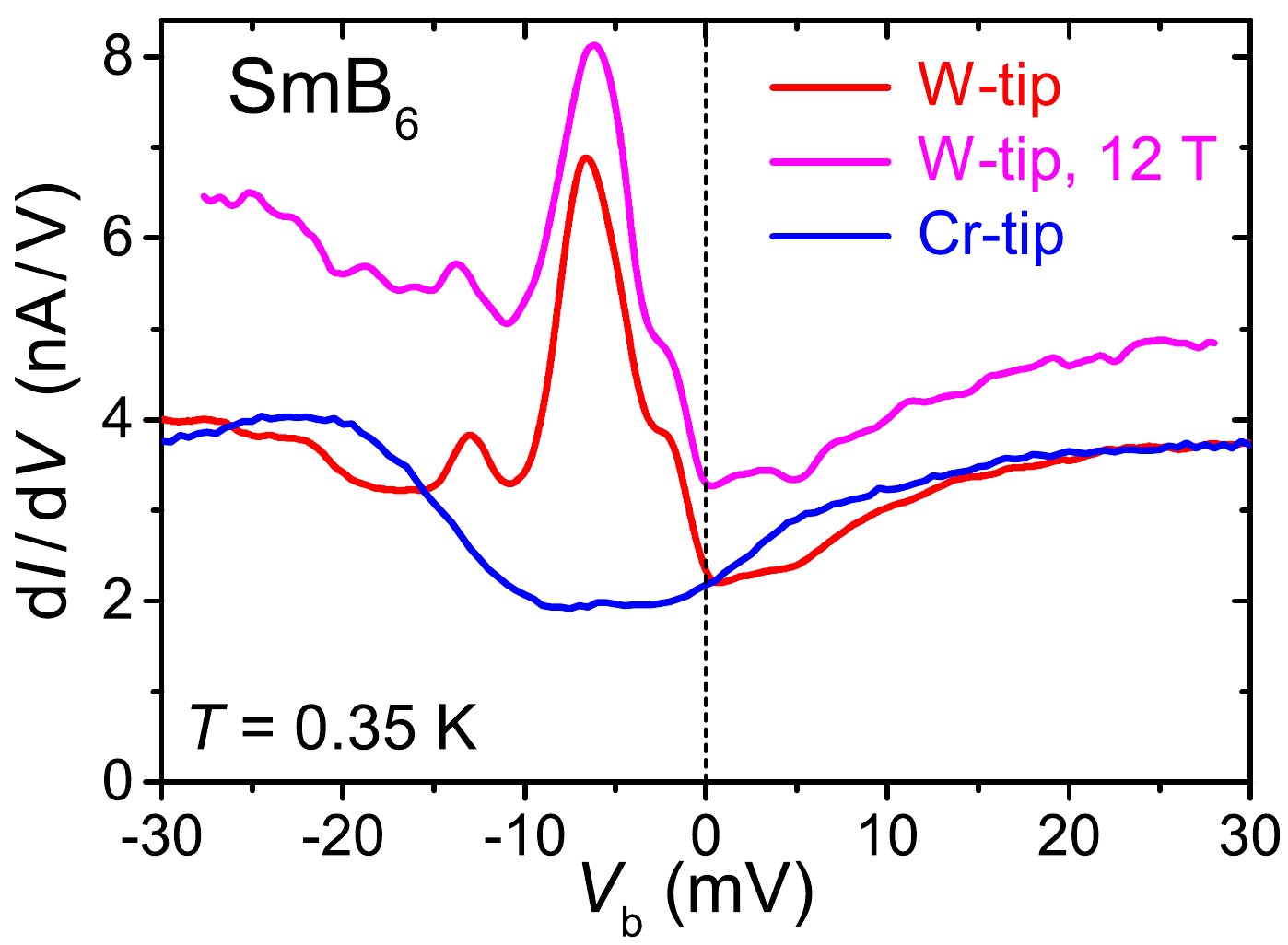}
\caption{\textbf{Tunneling spectra with W and Cr tips.}
Spectra obtained on non-reconstructed surfaces of pure SmB$_6$ by W-tip
(red) and magnetic Cr-tip (blue) at 0.35 K and zero magnetic field ($V_b$
= 50 mV, set-point current $I_{sp}$ = 200 pA). For comparison, a spectrum
taken with a W-tip at a magnetic field of 12 T is presented (pink,
vertically offset by 1 nA/V).} \label{spinpol}
\end{figure}

\begin{figure*}
\centering
\includegraphics[width=16.0cm]{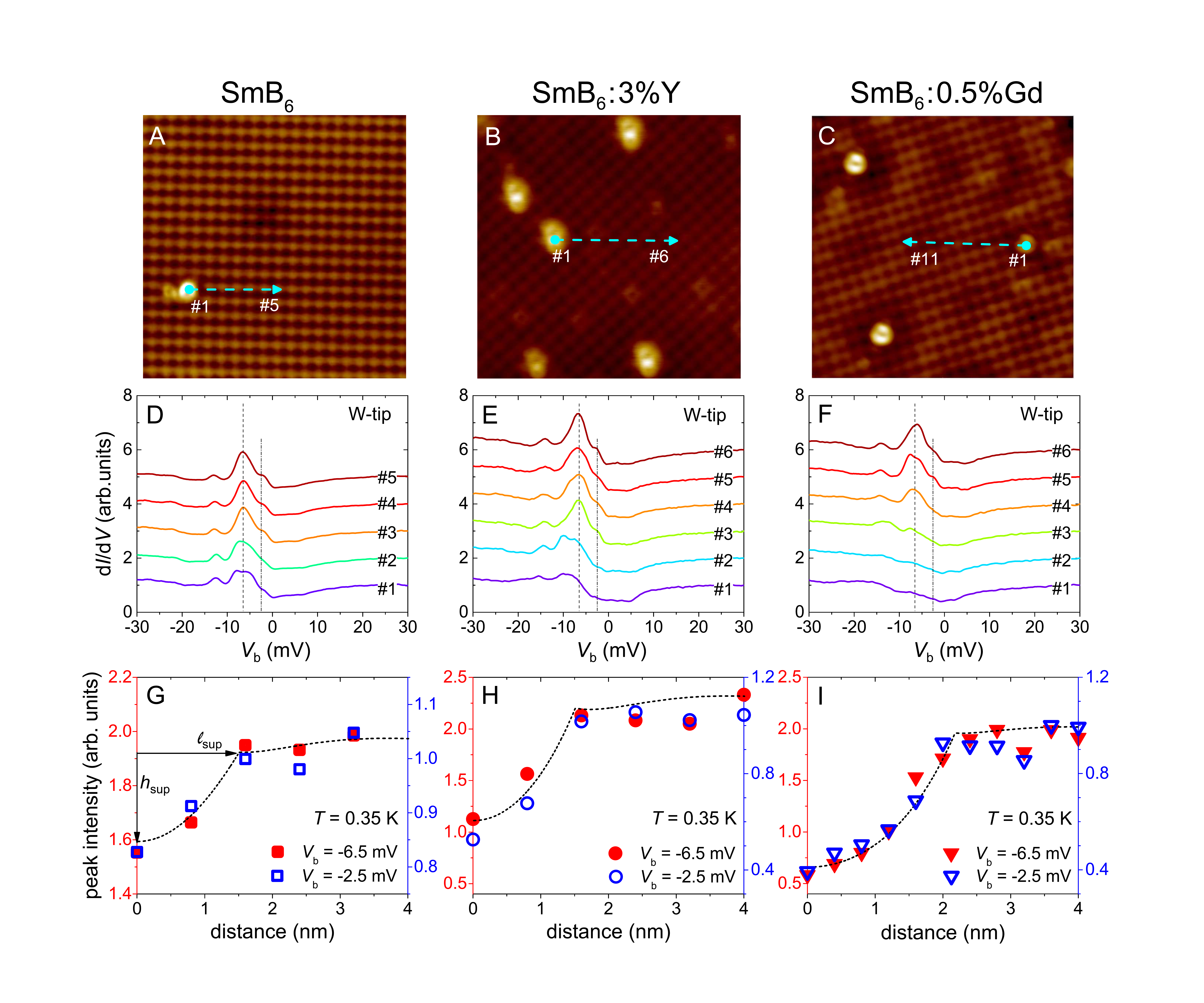}
\caption{\textbf{Influence of impurities on spectroscopic results.}
({\bf A})--({\bf C}) $8\! \times \! 8$ nm$^2$ topographies of pure as well
as 3\% Y- and 0.5\% Gd-substituted SmB$_6$. The cyan arrows indicate the
ranges and directions of STS measurements around the impurities.
({\bf D})--({\bf F}) d$I$/d$V$-curves of the three samples measured at
0.35 K and zero field. The curves are measured at positions with increasing
distance from the impurity (the impurities are located at \#1) along the
arrows in (A)--(C), correspondingly (bias voltage $V_b$ = 30 mV, current
set-point $I_{sp}$ = 100 pA). ({\bf G})--({\bf I}) d$I$/d$V$ values at
$V_b = -6.5$ meV (red) and $-2.5$ meV (blue) with increasing distance from
the impurity (impurities are located at 0). The black dashed lines are fits
according to the model described in \cite{suppl}. $h_{\rm sup}$ and
$\ell_{\rm sup}$ indicate the suppression of peak intensity at the impurity
and its lateral extent, respectively.} \label{pureYGd}
\end{figure*}

To investigate the $-6.5$ mV signature peak of the surface state in more
detail, we compare STS spectra obtained by a regular (non-magnetic) W-tip
and a magnetic Cr-coated tip (see \cite{suppl} for tip details) on the same
surface of pristine SmB$_6$. Cr-coated tips are often used for spin-dependent
STS \cite{bod03,oka14}. Figure \ref{spinpol} presents the tunneling spectra
at 0.35 K on a non-reconstructed surface (see Fig.\ \ref{pureYGd}A). At
large $|V_b| \gtrsim$~20 meV, the spectra measured by the two tips are very
similar and featureless indicating that any difference is not due to an
exotic DOS of the tips \cite{schl10}. For small $|V_b| \lesssim$ 20 meV,
however, the two d$I$/d$V$-spectra are markedly different: In the case of
the Cr-tip, the pronounced signature peak at $-6.5$ meV, and hence tunneling
into the surface state, is dramatically suppressed \cite{suppl} and the
direct bulk hybridization gap---albeit slightly reduced in size---is exposed \cite{yee13,ruan14,jiao16}. This is corroborated by a striking similarity of
spectra obtained with magnetic Cr-tip and such recorded with W-tip at 20~K
\cite{roe14,jiao16}, a temperature at which the surface state has not
formed. In addition, as we will show later, scanning with a W-tip over the
surface of Gd-substituted SmB$_6$ generates a similar reduction of the $-6.5$
meV peak at low temperatures. In this case, the W-tip may pick up magnetic Gd
substitutents from the surface, and this process can even be reversed (for
details see \cite{suppl}). Importantly, picking up Gd from the sample converts
a regular W-tip into a magnetic tip as, e.g., observed by STM on Fe$_{1+y}$Te
where excess Fe atoms were picked up \cite{ena14}. The close similarity of
the spectra obtained with these two types of magnetic tips suggests that the
spectral changes are induced by the magnetic nature of the tips, consistent
with a spin texture at the surface of SmB$_6$ \cite{bar16}. However, the
reduction in d$I$/d$V$ upon using magnetic tips, reaching 72\% at $V_{\rm b}
= -6.5$ mV, is, to the best of our knowledge, extraordinarily large and beyond
expectations for spin-polarized STS \cite{bod03,oka14}. Thus, spin-polarized
tunneling alone, based on an in-plane alignment of the Dirac electron spins,
may not account for this very effective suppression of the signature peak at
$-6.5$ meV. This is even more obvious in view of a spin polarization of less
than 50\% for a Cr-tip \cite{cor12}. Moreover, a tunneling spectrum obtained
at $\mu_0 H =$ 12~T is rather similar to zero-field spectra for regular W-tips
(Fig.\ \ref{spinpol}), and precludes the possibility of a magnetic stray field
of the magnetic tip suppressing the surface state locally.

To scrutinize the effect of magnetism on the surface states of SmB$_6$
we now investigate the \emph{local} impact of substituents, both non-magnetic
(Y) and magnetic (Gd), on these surface states. Figures \ref{pureYGd}A--C
exhibit representative topographies ($8 \times 8$ nm$^2$ field of view) of a
pristine, a 3\% Y- (SmB$_6$:3\%Y) and a 0.5\% Gd-substituted sample
(SmB$_6$:0.5\%Gd), respectively. By comparison to earlier work
\cite{roe14,roe16} on pristine SmB$_6$, we infer that these surfaces are
B-terminated. The surface of pure SmB$_6$ is very clean, Fig.~\ref{pureYGd}A,
exhibiting only very few defects. The protrusions seen in this topography are
likely non-magnetic \cite{jiao16}. Comparing the surfaces of the
substituted samples to pristine SmB$_6$ reveals the expected higher density
of defects in the former, and one can safely assume that the protrusions in
substituted samples predominantly represent the substituents \cite{suppl},
a fact that is also supported by the spectroscopic results below. Also, the
observation of W-tips changing into magnetic ones after they picked up atoms
(or clusters) from Gd-substituted SmB$_6$ surfaces suggests the involvement
of magnetic constituents, i.e.\ Gd, in the picked-up entities.

In the following, STS spectra were obtained on non-reconstructed, B-terminated
surfaces on which we focused on areas with only a very few defects. Figures
\ref{pureYGd}D--F represent spectra taken along the blue arrows shown in
the respective topographies A--C, i.e.\ spectra \#1 were taken on top of
the respective defects whereas spectra of increasing number were obtained
for increasing distance from the impurity. Clearly, the tunneling spectra
obtained at $T =$ 0.35 K sufficiently far away from the defects are strikingly
similar for all samples. This finding demonstrates that, within the current
substitution level, any influence of the defects is {\em highly local} in
nature, regardless of the magnetic properties of the substituent. However,
very close to---and specifically on top of---the substituents, there are
marked differences between the pure and Y-substituted SmB$_6$ on the one hand,
and the Gd-doped sample on the other hand: For non-magnetic defects, Figs.\
\ref{pureYGd}D and E, the surface state signature peak is only moderately
suppressed whereas in Gd-substituted SmB$_6$ all low-energy features appear
largely suppressed close to the magnetic substituent, and the spectrum is
reminiscent of those observed with Cr-tips, indicating a common origin of
the peak suppression.

To allow for a quantitative analysis of the impurity effect, the intensities
of the d$I$/d$V$-spectra at $-6.5$ meV and $-2.5$ meV as a function of the
distance from the defects are plotted in Figs.\ \ref{pureYGd}G--I. Note
that the additional shoulder at about $-$2.5 meV is {\em exclusively} related
\cite{jiao16} to the surface states (likely to the heavy quasiparticle
surface states \cite{luo15}). Both peaks (dotted lines in Fig.\
\ref{pureYGd}G--I) recover in a similar fashion upon going away from the
defects in these samples, but clearly $\ell_{\rm sup}$ and, specifically,
$h_{\rm sup}$ are quite different ($h_{\rm sup}$ and $\ell_{\rm sup}$
describe the peak intensity suppression at the defect and the extent of this
suppression, respectively). Peak intensities have regained their values on
unperturbed surfaces at $\ell_{\rm sup} \lesssim 1.5$ nm for pristine and
Y-substituted SmB$_6$, and $\ell_{\rm sup}^{\rm Gd} \approx 2.2$ nm for
Gd-substituted SmB$_6$. The recovery of the surface state with increasing
distance from the defect follows the prediction by theoretical models
\cite{liu09,wang10}. The dashed-line fits are described in \cite{suppl},
$h_{\rm sup}$ and $\ell_{\rm sup}$ are obtained from the respective curves.
On top of the non-magnetic defects, the TSS still survive. Although the
above-mentioned model is not intended to be applied to non-magnetic impurities
we made use of the fact that it describes the experimental data reasonably
well to still obtain $h_{\rm sup}$ and $\ell_{\rm sup}$. The apparent
applicability of the theoretical model to non-magnetic impurities along with
the moderate suppression ($h_{\rm sup} \approx$ 15\% -- 45\%) may be due to
the local changes of the bulk band structure \cite{liu09,wang10,bis10} and/or
the Kondo hole effect \cite{morr11,fuh17}. On the other hand, the large
magnetic moment of Gd locally breaks time reversal symmetry and may eventually
gap out the Dirac cone states. As a result, $h_{\rm sup}^{\rm Gd}$ reaches
70\% at the Gd defect, reminiscent of the value obtained by using a Cr-tip. A
similar influence of magnetic substituents on the tunneling spectra was also
observed in weakly correlated topological insulators, such as Cr-substituted
\cite{yan13,jia15,lee15} or V-substituted \cite{ses16} Sb$_2$Te$_3$. Our
observation of an only local impact of magnetic substituents on the surface
state is consistent with an unexpectedly insensitive response of the
TSS in Bi$_2$Se$_3$ to magnetic impurities at low impurity concentration in
a macroscopic measurement \cite{val12}.

As shown above, the TSS are fully recovered around 2.2 nm from the Gd
substituent site. For SmB$_6$:0.5\%Gd, the average distance between Gd
substituents is $\sim$2.4~nm. A pressing question at this juncture is:
what if the average Gd-Gd distance is reduced to well below $\ell_{\rm sup}$,
i.e.\ if the areas of suppressed surface states sufficiently overlap?
To address that, we also probed a SmB$_6$:3\%Gd sample with average Gd-Gd
distance of about 1.3~nm. Here, a surface state signature peak was only found
in areas where the statistical distribution of Gd atoms resulted in larger
distances between them.

\begin{figure}
\centering
\includegraphics[width=8.2cm]{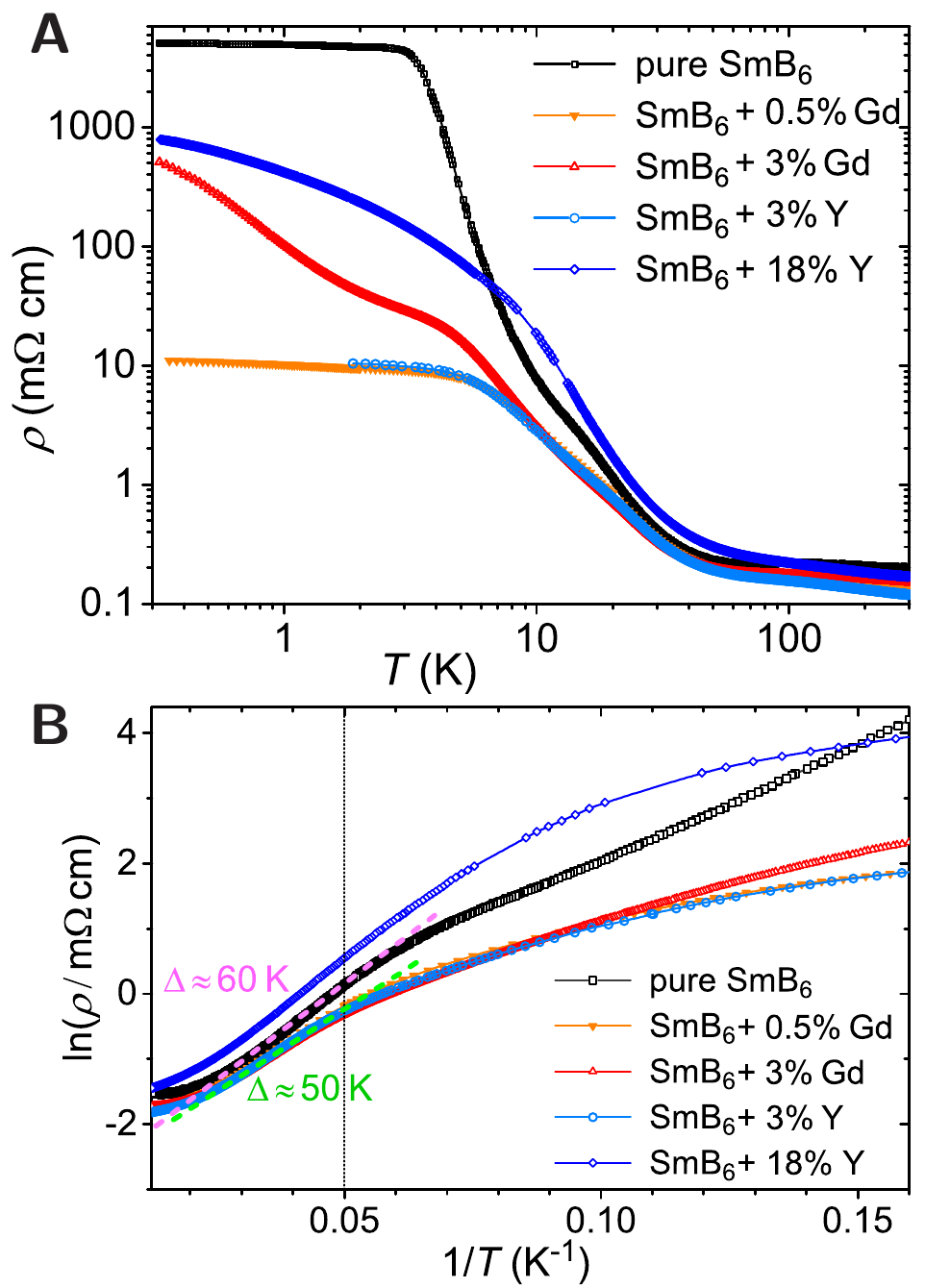}
\caption{\textbf{Resistivity of pristine and substituted SmB$_6$.}
({\bf A}) Temperature dependence of resistivity $\rho$ pure and differently
substituted SmB$_6$ in double-logarithmic presentation. ({\bf B}) $\ln (\rho)$
versus 1/$T$ plot at intermediate temperatures used to derive the energy
gap from thermal excitation. The gap values obtained from the slopes of the
pristine (pink dashed line) and the lightly substituted samples (green
dashed line) above 20 K (dotted vertical line) are given.} \label{resist}
\end{figure}

The resistivity $\rho(T)$ of pure SmB$_6$ exhibits a well-known saturation
below around 3 K which is due to surface conductance \cite{wol13,kim13},
see Fig.\ \ref{resist}A. A very similar behavior is found for SmB$_6$:3\%Y
and SmB$_6$:0.5\%Gd samples, yet with a much smaller overall change in
$\rho(T)$ due to the substituents. For SmB$_6$:3\%Gd, the low-$T$ saturation
of $\rho(T)$ is indeed not observed, instead $\rho(T)$ continues to increase
exponentially, indicating a remaining gap. This is expected when the
average Gd-Gd distance is smaller compared to $\ell_{\rm sup}^{\rm Gd}$.
In highly, non-magnetic substituted SmB$_6$, see example of 18\% Y in Fig.\
\ref{resist}, the $\rho(T)$-behavior is qualitatively different from non-
or lightly substituted samples, possibly due to interacting substitutents.

The data presented in Fig.~\ref{resist}B allow an estimate of the changes
exerted on the bulk hybridization gap $\Delta$ from $\rho(T) \propto
\exp(\Delta/k_B T)$. Pure SmB$_6$ exhibits the typical two gap values
\cite{ris00} with $\Delta_1 \approx$ 36~K for 5~K $\leq T \leq$ 12~K and
$\Delta_2 \approx$ 60~K for 20~K $\leq T \leq$ 40~K (the latter is marked
in Fig.\ \ref{resist}B). For the lightly ($\leq 3$\%) substituted samples,
somewhat reduced gap values \cite{ore17} of $\Delta_1 \approx$ 24~K
(9~K $\leq T \leq$ 14~K) and $\Delta_2 \approx$ 50~K at higher $T$ are
observed, along with an increased surface conductivity, all in line with an
substitution-induced modification of the Kondo lattice formation \cite{kim14}.
Yet, these changes in the bulk are minute and apparently too small to
account for the dramatic changes in the surface properties. Above $\sim$10~K,
the resistivities are determined by the bulk band structure, and the measured
values perfectly overlap for the lightly substituted samples. In contrast,
$\rho(T)$ of the highly substituted sample SmB$_6$:18\%Y below $\sim$20~K
deviates from exponential behavior.

The resistivity data in Fig.\ \ref{resist} provide compelling support from
a global measurement for the local picture obtained from STS (Fig.\
\ref{pureYGd}): Around a magnetic substituent, the disturbance is stronger
and extends further out compared to non-magnetic impurities. In the former
case, the formation of a {\em global} conducting surface state in
SmB$_6$:3\%Gd at low $T$ is already inhibited, providing a microscopic
picture of how the topologically protected surface state is destroyed {\em in
real space}.

Now the pressing question concerns the underlying mechanism for the
suppression of the surface state signature peak in STS in both cases, for
magnetic tips as well as magnetic substituents in SmB$_6$. The observed
disappearance of the peaks at $-6.5$ meV and $-2.5$ meV upon tunneling with
magnetic tips or on surfaces of Gd-substituted samples could be either due to
a suppression of the actual surface states, or by simply suppressing the
tunneling probability into the corresponding states (or a combination
thereof). Although we cannot unambiguously distinguish between these two
scenarios we consider the similarity of the spectra with suppressed surface
state signature peak to those obtained on pristine SmB$_6$ with W-tip
\cite{jiao16} at $T = 20$ K, i.e.\ a temperature at which the surface
states have not yet formed, as a strong indication towards the former,
i.e.\ a repressed formation of the surface state, see Fig.\ S4. The main
parameter determining the extent of the suppressed surface state around a
magnetic impurity is related to the exchange interaction \cite{liu09}.
Moreover, the surface state suppression by using magnetic tips calls for an
interaction whose energy scale is well beyond the Zeeman energy scale
associated with a magnetic field of 12~T, Fig.\ \ref{spinpol}. Therefore,
we propose an exchange-interaction based proximity effect to be involved
when tunneling with a magnetic tip or around a magnetic substituent.

Our findings have two important consequences. First, they provide a
microscopic picture of how the surface states are perturbed by impurities.
This pertubation takes place locally at the defect site, with an extent
$\ell_{\rm sup}$ that depends on the magnetic properties of the defect.
Enhanced values of $\ell_{\rm sup}$ and, particularly, $h_{\rm sup}$ at
magnetic substituents as observed by our STM experiments were considered
a hallmark for TSS \cite{liu09}. Secondly, the very effective suppression
of the surface state signature peak at $-6.5$ meV can be exploited in
applications. We propose to use SmB$_6$ to detect exchange fields. If a
tunneling tip is made of SmB$_6$ and scanned over a surface to be
investigated, the d$I$/d$V$-response at $V_{\rm b} = -6.5$~mV is expected
to change significantly around a magnetic surface atom. Based on our
investigations by magnetic tips and on magnetic impurities this effect
should allow for single spin detection.

\paragraph* {Acknowledgments:}
We acknowledge valuable discussion with Po-Yao Chang, Piers Coleman, Onur
Erten, Mohammad H. Hamidian, Dae Jong Kim, Michael Nicklas, Dirk Sander
and Liu Hao Tjeng. This work was supported by the DFG through SPP 1666.
L.J. acknowledges support by the Alexander-von-Humboldt foundation. P.F.S.R.
acknowledges support from the Laboratory Directed Research and Development
program of Los Alamos National Laboratory under project number
20160085DR.\\[0.6cm]

\newpage \clearpage
\renewcommand\figurename{FIG. S$\!$}
\setcounter{figure}{0}
\section*{Supplementary materials}

\subsection*{I. Sample preparation}
All samples used in this study were grown by the Al-flux method \textit{(9)}.
Single crystals were cleaved \textit{in situ} below 20 K to expose a (001)
surface. For pristine SmB$_6$, 6 different single crystals were cleaved
and investigated for this study, for 0.5 at.\% Gd-substituted SmB$_6$ and
3 at.\% Y-substituted SmB$_6$ three single crystals were investigated.

Pristine SmB$_6$ sample are difficult to cleave and atomically flat and
well resolved surface areas have to be searched for. By introducing
substituents into SmB$_6$, the cleavage properties change dramatically and
atomically flat areas can be found much more easily. However, the vast
majority of the surface areas investigated so far was reconstructed (see
also Fig.\ S\ref{Gdrec} and related discussion). Again, unreconstructed
surface areas have to be searched for.

\subsection*{II. Details of Tunneling measurements}
STM measurements were conducted in an ultra-high vacuum ($p < 3 \cdot
10^{-9}$ Pa) environment and at a temperature $T =$ 0.35 K. The tunneling
current $I$ was measured using tungsten tips or Cr-coated tips. Tunneling
parameters for topography, if not noted otherwise, were $V_b =$ 300 mV and
$I_{sp} =$ 200 pA. The differential conductance (d$I$/d$V$) spectra were
acquired by lock-in technique applying a modulation voltage of typically
$V_{\rm mod} =$ 0.3~mV; if enhanced resolution was strived for, $V_{\rm mod}$
was reduced to 0.05 mV. The bias voltage $V_b$ is applied to the sample. A
magnetic field of up to 12 T can be applied perpendicular to the scanned
sample surface.

\subsection*{III. Scanning Tunneling Spectroscopy using Cr-tips}
For spin-polarized scanning tunneling spectroscopy commercially available
\begin{figure}[t]  \centering
\includegraphics[width=7.8cm]{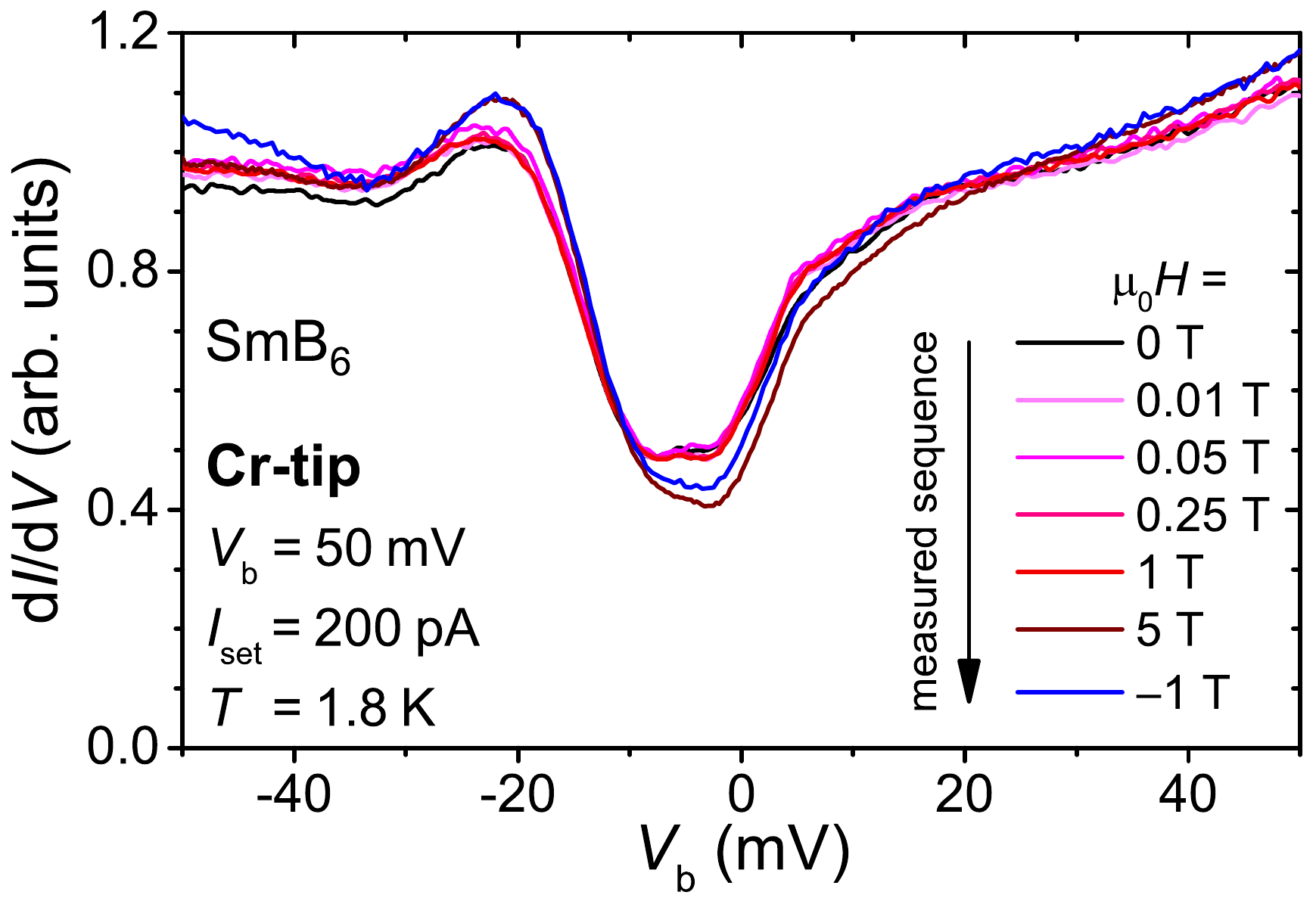}
\caption{STS with Cr-tip in magnetic field. Spectra were obtained with a
Cr-coated tip on pristine SmB$_6$. The magnetic field was successively
increased to 5 T, then decreased to zero field, and increased to $-1$ T in
reversed direction. As for Cr-coated tip in zero field, the sharp peak at
$V_b \simeq -6.5$ mV indicative of the surface state in SmB$_6$ is strongly
suppressed in magnetic fields, exposing the Kondo gap. The STS data exhibit
only a small sensitivity to magnetic fields indicating the robustness of the
observed peak suppression.} \label{Cr-field} \end{figure}
Cr-coated tips (NaugaNeedles LLC:
http://nauganeedles.com/products-USSTM-W500-Cr) were used. Such tips are
characterized by uncompensated magnetic moments at the Cr tip apex resulting
in a spin-polarization (up to 45\%) at the Fermi level \textit{(30)}.
In addition to STS on SmB$_6$ with magnetic tips in zero field, such
measurements have also been conducted in magnetic fields. Selected results
of one of the field cycles are presented in Fig.\ S\ref{Cr-field}. Here, the
magnetic field (applied perpendicular to the sample surface) was gradually
increased up to $\mu_0 H = 5$ T, consecutively ramped back down to zero field
and reversed, with spectra taken at constant field values. Clearly, no
significant change in the tunneling spectra is observed. The d$I$/d$V$-data
perfectly overlap in the low-field regime, i.e.\ weak-antilocalization (WAL)
effects are not visible in our tunneling spectra. At high magnetic fields
the zero-bias conductance is slightly reduced. Such a high applied field may
influence the magnetization orientation within our magnetic tip which, in
turn, can reduce the tunneling current through spin-polarized effects.

Notably, scanning the Cr-tip over substantial surface areas of pristine
SmB$_6$ alludes towards the lack of any significant local dependencies of the
spectra.

\subsection*{IV. Changing tip conditions on Gd-substituted SmB$_6$}
\begin{figure}[t]\centering
\includegraphics[width=8cm]{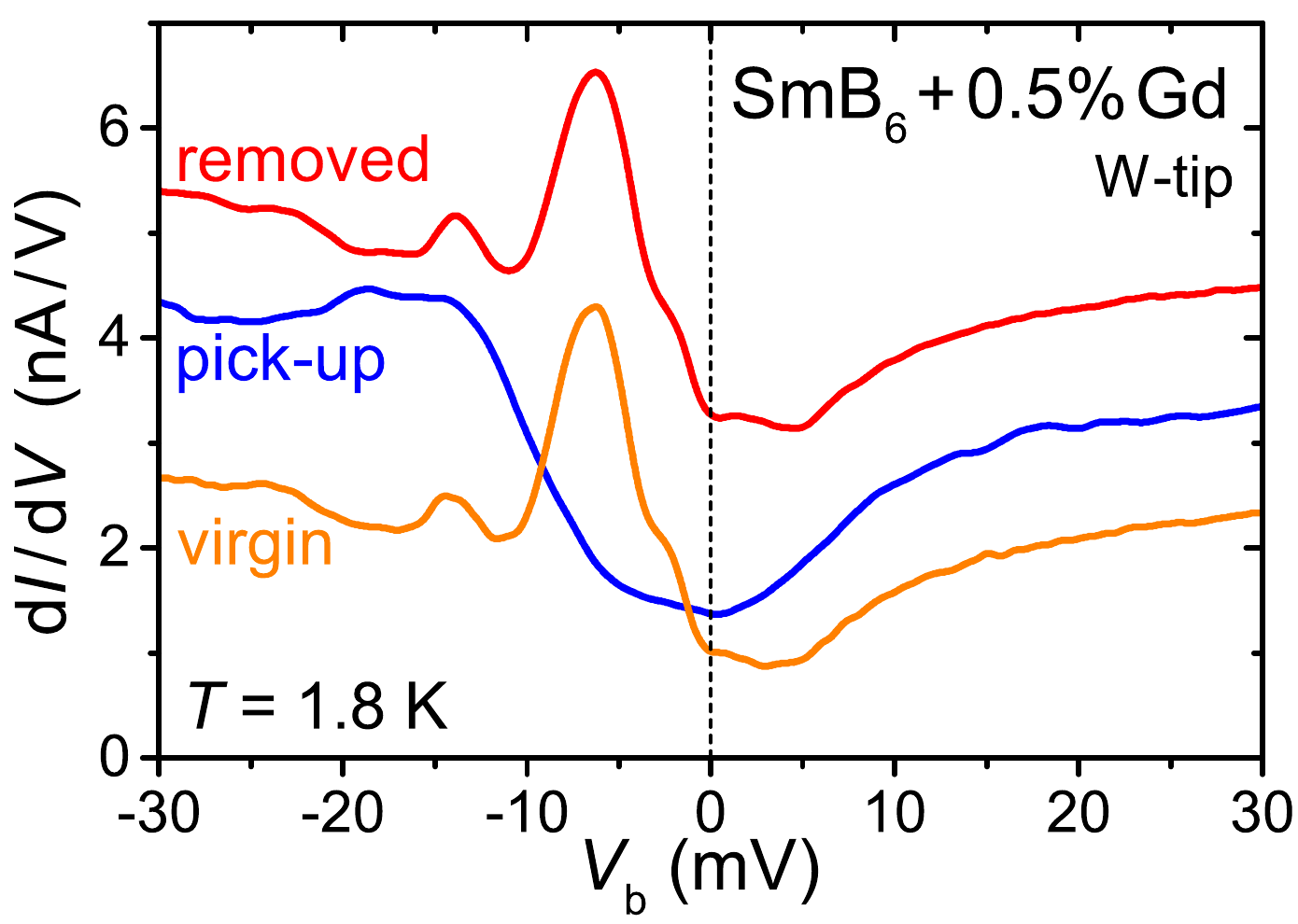}
\caption{Converting a non-magnetic into a magnetic tip. All Spectra were
obtained on the same non-reconstructed surface of 0.5\% Gd-substituted
SmB$_6$. Virgin W-tips showed the orange spectrum (virgin), but changed
after some scanning into the blue one (pick-up), likely due to picked-up Gd
atoms. After applying voltage pulses to the tip, the original spectrum could
be restored (red, removed). The blue spectrum matches nicely with those
obtained with Cr-tips (Fig.\ 1 of main text, blue spectrum)
indicating that the tip is converted into a magnetic one by picking up
entities from the sample surface.} \label{Gdpick}
\end{figure}

In the following we discuss observations on 0.5 at.\% Gd-substituted SmB$_6$,
investigated using W-tips. Initially, the STS spectra obtained on surfaces of
such samples and away from defects were very similar to those taken on pure
SmB$_6$. One example is presented in Fig.\ S\ref{Gdpick}, orange line marked
as ``virgin" (SmB$_6$:0.5Gd, for topography see Fig.\ 2C of the main text).
In particular, the surface state signature peak at $V_{\rm b} \approx -6.5$ mV
is well developed at $T =$ 1.8 K.

\begin{figure}[t]  \centering
\includegraphics[width=8.4cm]{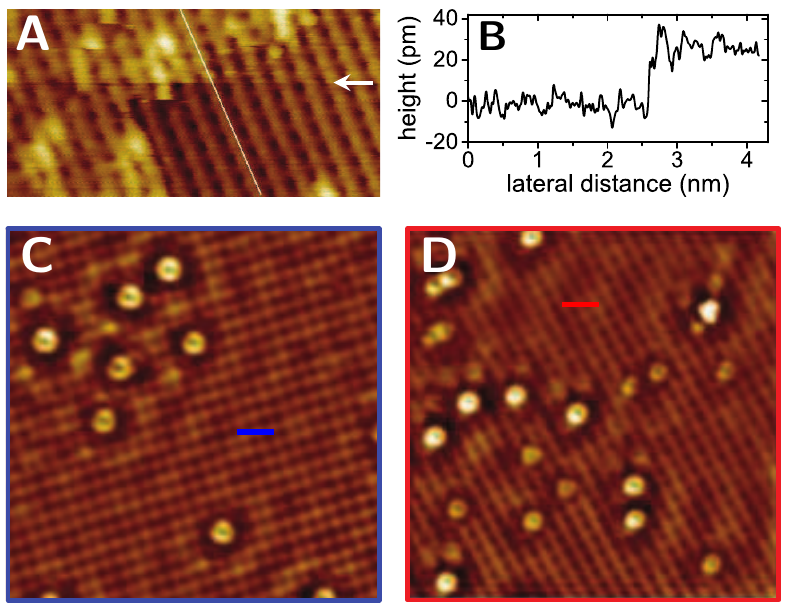}
\caption{Topographies obtained with a W-tip on 0.5\% Gd-substituted SmB$_6$
showing the changing tip condicitons. (A) While scanning the tip often changed
suddenly. Here, such an individual tip change is marked by a white arrow (area
$8 \times 4$ nm$^2$, up-scan). (B) Height profile along the white line in (A).
The upward step is likely related to the tip having picked up something from
the surface. (C) Topography ($10 \times 10$ nm$^2$) after extensive scanning
on the surface and numerous tip changes as in (A). Blue line indicates a 1 nm
line away from defects over which spectra were taken and averaged to yield the
blue curve of Fig.\ S\ref{Gdpick}. (D) Topography ($10 \times 10$ nm$^2$)
after applying several voltages pulses (up to $+10$ V) to the tip. Red line:
line over which spectra were averaged for red curve in Fig.\ S\ref{Gdpick}.
Because of the voltage pulses different areas were scanned.}
\label{pick} \end{figure}

In Fig.\ S\ref{pick} we present more examples of topographies obtained on a
surface of a 0.5\% Gd-substituted SmB$_6$ sample. Atomically resolved
topographies are difficult to obtain on surfaces of substituted SmB$_6$
(specifically for higher substitution levels) as the tips are frequently
changing while scanning, likely due to picked-up atoms or clusters. Note
that such a susceptibility to picking up atoms from the surface is in
line with the fact that the substituted samples cleave much more easily
compared to pure SmB$_6$. A sudden, individual tip change is shown in the
upward scan Fig.\ S\ref{pick}A and marked by a white arrow. The effective
height difference across this step, taken along the white line in Fig.\
S\ref{pick}A and plotted in Fig.\ S\ref{pick}B, of about 25 pm is not expected
from the crystallographic structure of SmB$_6$, and no downward step is
observed at this position in the subsequent down-scan. Hence, we argue that
the tip picked up something from the surface at this particular scan line,
consistent with a decreasing tip-sample distance. Since the vast majority of
the defects seen in the topographies of substituted samples is caused by the
substituents it is highly likely that the tip picked up Gd in this case.

After extensive scanning and several changes in topographic height as just
described, we sometimes observed topographies like the one presented in Fig.\
S\ref{pick}C. Apparently, some tips which are modified by picking up some
Gd atoms (or Gd-containing clusters) from the sample surface are reasonably
stable to produce atomically resolved topographies. The blue line (length of
1 nm) indicates the line along which 101 spectra were taken and averaged to
give the blue curve (``pick-up'') of Fig.\ S\ref{Gdpick}. Clearly, this
spectrum is extremely similar to the one obtained with Cr-coated tip, Fig.\ 1
of the main text. We note that these spectra were intentionally obtained away
from any (visible) defect on the surface. Only after applying several voltage
pulses (typically of +10 V) to the tip, these adatoms could be removed from
the tip. A topography after such a removal of adatoms is presented in
Fig.\ S\ref{pick}D. We note that such severe voltage pulses significantly
disturb the sample surface close to the tip and hence, different sample areas
were investigated before and after applying voltage pulses to the tip. The
thereafter obtained spectrum, red curve in Fig.\ S\ref{Gdpick} (``removed'')
taken along the red line in Fig.\ S\ref{pick}D, corresponds to the virgin one.

The spectra obtained with Cr-coated tip and with W-tip after picking up atoms
(or clusters) from the surface (blue curve in Fig.\ S\ref{Gdpick}) exhibit a
particularly effective suppression of the surface state signature peak. Along
with the above-mentioned fact that Gd was likely picked up, we surmise that
the changes in spectra are due to a conversion of the W-tip apex into a
magnetic one by picking up Gd while scanning, similar to observations on
Fe$_{1+y}$Te \textit{(28)} by picking up excess Fe. Extensive scanning revealed
that such spectra with suppressed surface state signature peak were obtained
everywhere on the sample surface, and not just by positioning the tip
accidently on top of a Gd-substituent. The fact that we can reverse this
process through cleaning the tip by applying voltage pulses, i.e.\ turning
the W-tip back to a normal (non-magnetic) one, heavily supports this
\begin{figure}[t!]  \centering
\includegraphics[width=8.2cm]{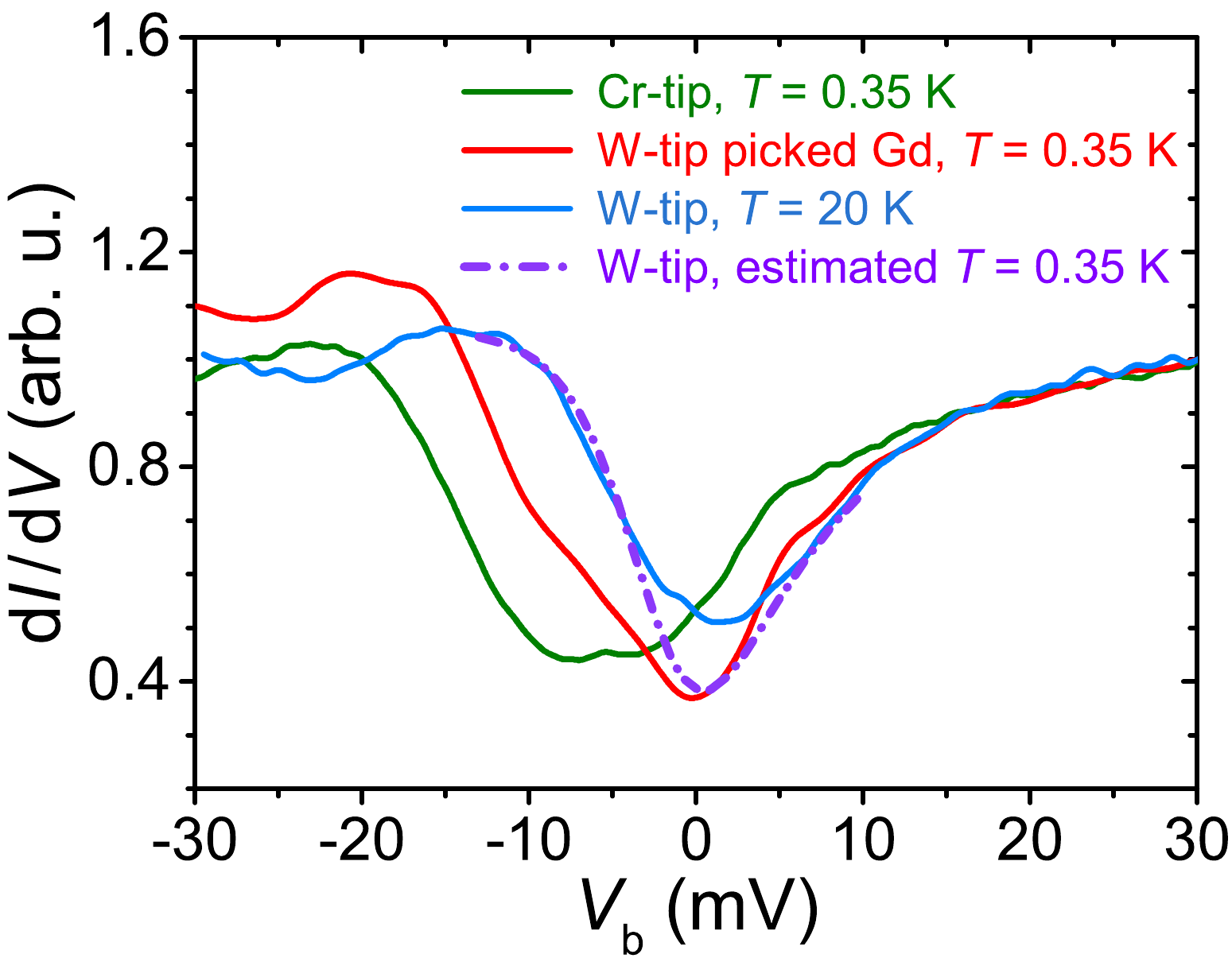}
\caption{The surface state signature peak of SmB$_6$ at $V_b = -6.5$~mV can
be suppressed by using magnetic tips (Cr-coated or after picking up Gd from
the sample surface, green and red curve, respectively) or by using regular
W-tips and raising the temperature to 20 K. From the latter, a d$I$/d$V$-curve
at 0.35~K is estimated (violet dash-dotted line) which coincides with the
20~K-data after thermal broadening.}
\label{comp} \end{figure}
conjecture. We emphasize that such changes to the tip, i.e.\ picking up atoms
or clusters from the surface with the concomitant spin-polarization in the
spectra, and recovering the regular spectra after applying voltage pulses to
the tip, were conducted repeatedly on several surfaces of samples with
different Gd substitution level and with different W-tips, all with consistent
results. Such tip changes were {\em not} observed on pristine SmB$_6$ (for
pure SmB$_6$ more than 30 cleaved surfaces were investigated so far). This
indicates that Gd is important for observing these changes in the obtained
spectra; its magnetic nature provides a reasonable explanation for observing
spectra very similar to the ones seen with spin-polarized Cr-tips. Our
corresponding results not only support the spin-polarized nature of tunneling
with picked-up tips but also the assignment of the defects observed on the
surfaces of Gd-substituted samples to Gd impurities.

\subsection*{V. Comparison of spectra with suppressed surface state
signature peak}
We have presented several ways to suppress the peak in the tunneling spectra
at $V_b = -6.5$~mV which signals the surface state in pure SmB$_6$. This peak
suppression can be achieved by using a magnetic tip (either Cr-coated or after
\begin{figure}[t]  \centering
\includegraphics[width=6.0cm]{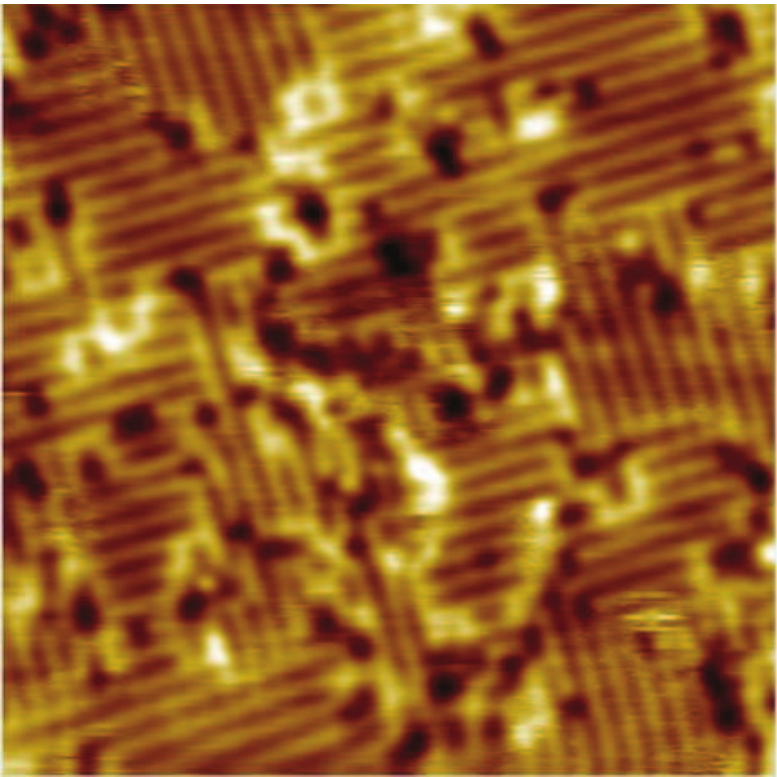}
\caption{Disordered reconstructed surface of a 0.5\% Gd-substituted SmB$_6$
sample, field of view $20 \times 20$ nm$^2$, $V_{\rm b} =$ 0.1 V,
$I_{\rm sp} =$ 0.4 nA, $T =$ 1.8 K.}
\label{Gdrec} \end{figure}
having picked up Gd by a W-tip from a Gd-substituted sample) or locally around
defects. In Fig.\ S\ref{comp} we compare such spectra obtained with magnetic
tips, to a d$I$/d$V$-spectrum obtained at 20~K on a pure SmB$_6$ sample
\textit{(8)}, blue curve in Fig.\ S\ref{comp}. At this temperature, the
surface states have not yet formed. In order to allow for a comparison of the
different temperatures, we estimated a curve at $T =$ 0.35~K (dash-dotted
curve in Fig.\ S\ref{comp}) which, thermally broadened to $T =$ 20~K,
coincides with the spectrum measured at 20~K. At $T =$ 0.35~K, all spectra
exhibit a similar suppression of the tunneling conductance d$I$/d$V$ at low
$V_b$, thereby exposing the bulk Kondo gap. This comparison indicates that
very likely the surface states themselves are suppressed in all cases, rather
than the tunneling probability into the surface states.

\subsection*{VI. Reconstructed surface of Gd-substituted SmB$_6$}
Pristine SmB$_6$ is notoriously difficult to cleave. This changes immediately
upon substitution of Sm by other rare earths. Nonetheless, the complexity of
the surface terminations as observed for pristine SmB$_6$ \textit{(22)}
remains in the substituted samples. In particular, large portions of the
surfaces of substituted SmB$_6$ are reconstructed. An exemplary topography
of a reconstructed surface of 0.5\% Gd-substituted SmB$_6$ is shown in Fig.\
S\ref{Gdrec}. Very likely this results, as in the case of pristine SmB$_6$,
from the polar nature of the $\{$0~0~1$\}$ surfaces investigated here.

\subsection*{VII. Defect analysis of the 3\% Y-substituted SmB$_6$}
\begin{figure}[t]  \centering
\includegraphics[width=7.2cm]{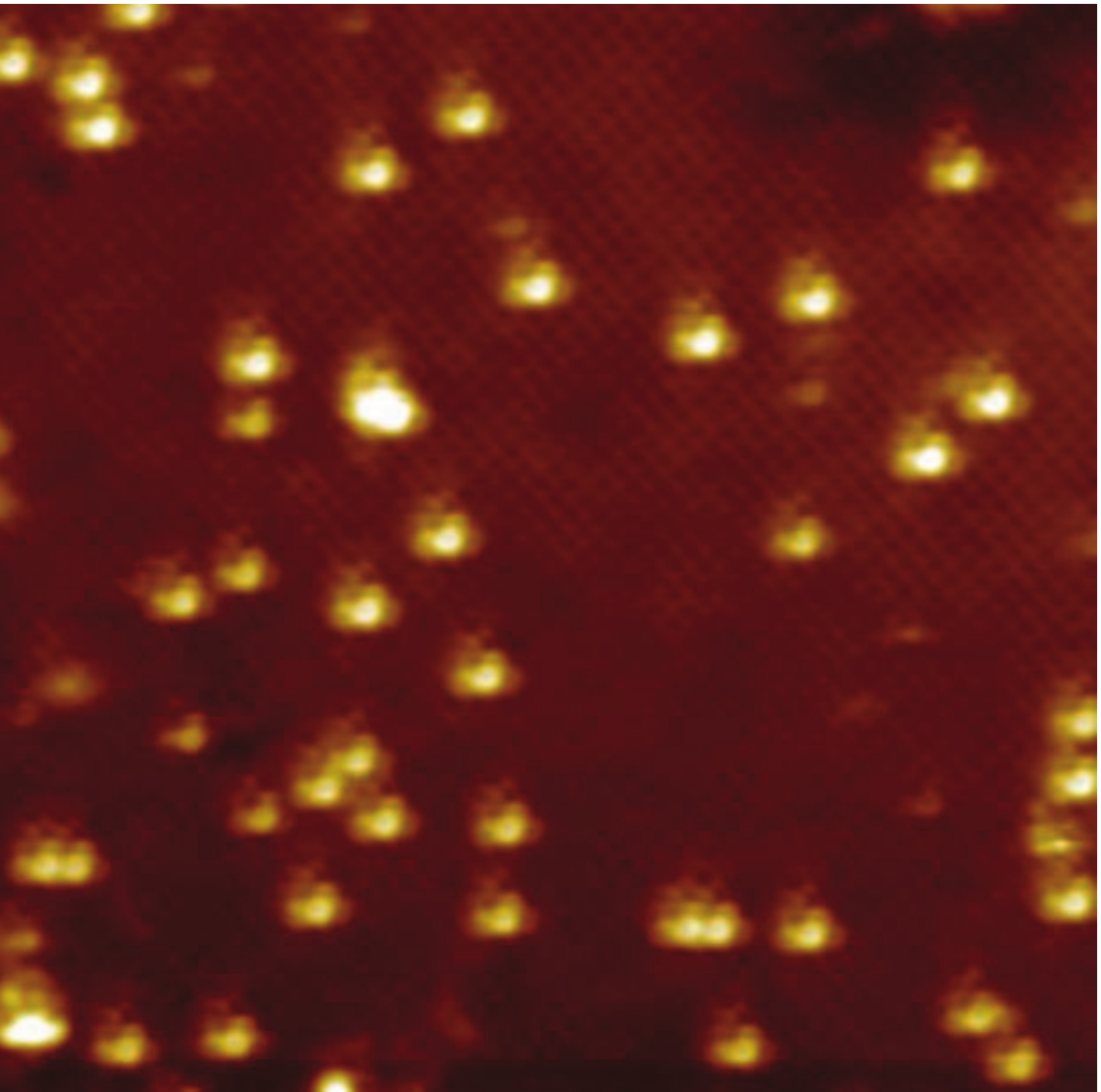}
\caption{Surface of 3\% Y-substituted SmB$_6$. Within the area presented
($20 \times 20$ nm$^2$) about 50 defects can be counted. The atomically
resolved surface is likely B-terminated.}
\label{3Ytopo} \end{figure}

In Fig.\ S\ref{3Ytopo} a representative surface of a  3\% Y-substituted sample
is presented. Within the shown area of $20\times 20$ nm$^2$ approximately 50
defects can be counted which is slightly less than expected if all defects
are caused by Y substituents. Similar defect densities were observed within
other areas on the same surface. We note that reconstructed areas were not
analyzed as in these cases the defects are considerably more difficult to
assess, see e.g. the Gd-substituted sample of Fig.\ S\ref{Gdrec}.

\subsection* {VIII. Analytical solution to the single magnetic impurity}
In this section, we will theoretically discuss the LDOS in a model describing
the Dirac surface state of a topological insulator coupled to a single
magnetic impurity with exchange interaction [Ref.\ 32,  A.\ Matulis, F.\ M.\ 
Peeters, Quasibound states of quantum dots in single and bilayer graphene. 
{\it Phys. Rev. B} \textbf{77}, 115423 (2008)]. The model
Hamiltonian of such coupling in polar coordinate is given by
\begin{equation}\nonumber
H_0 = \left(
      \begin{array}{cc}
      M_0\Theta(r_0-r) & -iAe^{i\theta}(\frac{\partial}{\partial r}+\frac{i}{r}\frac{\partial}{\partial \theta})\\
      -iAe^{-i\theta}(\frac{\partial}{\partial r}-\frac{i}{r}\frac{\partial}
      {\partial \theta}) & -M_0\Theta(r_0-r)\\
      \end{array}
      \right),
\label{eq1}
\end{equation}
where $A$($\equiv$ $\hbar v_F$) is the Fermi velocity of the Dirac electrons,
$M_0$ = $J_zS_z$ is the exchange interaction between Dirac electrons and the
magnetic impurity, $r_0$ determines the range of the exchange interaction,
and $\Theta$($r_0$-$r$) is the step function. As we will see below, $r_0$ will
determine the suppression range $\ell_{\rm sup}$ while $M_0$ will determine
the maximal suppression ratio $h_{\rm sup}$ of LDOS measured in experiments.
Due to the rotation symmetry of the above Hamiltonian, the wave function will
take the ansatz
\begin{equation}\nonumber
\Psi(n) = \frac{1}{\sqrt{2\pi}}\left(
      \begin{array}{cc}
     e^{in\theta}\phi_1(r)\\
     e^{i(n-1)\theta}\phi_2(r)\\
      \end{array}
      \right),
\label{eq2}
\end{equation}
where $n$ labels the angular momentum and is a good quantum number. Based on
this wave function ansatz, the Schr\"{o}dinger equation can be simplified as
\begin{equation}\nonumber
\left(
      \begin{array}{cc}
      M_0\Theta(r_0-r) & -iA(\frac{\partial}{\partial r}-\frac{n-1}{r})\\
      -iA(\frac{\partial}{\partial r}+\frac{n}{r}) & -M_0\Theta(r_0-r)
      \end{array}
\right)\!\!
\left(
      \begin{array}{cc}
      \phi_{1,n}\\
      \phi_{2,n}
      \end{array}\!
\right) = E \! \left(
      \begin{array}{cc}
      \phi_{1,n}\\
      \phi_{2,n}
      \end{array}\!
\right)\! .
\label{eq3}
\end{equation}
The resulting equation for $\phi_1$ is given by
\begin{eqnarray}\nonumber
&\frac{d^2}{dr^2}\phi_{1,n}(r)+\frac{1}{r}\frac{d}{dr}\phi_{1,n}(r)\\
&+(\frac{E^2-M_0^2\Theta(r_0-r)}{A^2}-\frac{n^2}{r^2})\phi_{1,n}(r) = 0.
\label{eq4}
\end{eqnarray}
Here, $E$ denotes the energy difference between the STS bias voltage and
the Dirac point.

The equation for $\phi_{1,n}$ takes the general form of (modified) Bessel
functions and thus, we can further simplify our solution. When $|E| \geq
|M_0|$, our solution ansatz will be
\begin{equation}\nonumber
\begin{aligned}
\phi_n = &\frac{1}{N}\left(
      \begin{array}{cc}
     a(M_0+E)J_n(k_1r)\\
     -iAk_1aJ_{n-1}(k_1r)
      \end{array}
      \right)\qquad\qquad\qquad\quad r<r_0,\\
     &\frac{1}{N}\left(
      \begin{array}{cc}
    E(bJ_n(k_2r)+cY_n(k_2r))\\
    -iAk_2(bJ_{n-1}(k_2r)+cY_{n-1}(k_2r))
      \end{array}
      \right)\quad r>r_0,
\end{aligned}
\label{eq13}
\end{equation}
and when $|E|$ $<$ $|M_0|$,
\begin{equation}\nonumber
\begin{aligned}
\phi_n = &\frac{1}{N}\left(
      \begin{array}{cc}
     a(M_0+E)I_n(k_1r)\\
     -iAk_1aI_{n-1}(k_1r)
      \end{array}
      \right)\qquad\qquad\qquad\quad r<r_0,\\
     &\frac{1}{N}\left(
      \begin{array}{cc}
    E(bJ_n(k_2r)+cY_n(k_2r))\\
    -iAk_2(bJ_{n-1}(k_2r)+cY_{n-1}(k_2r))
      \end{array}
      \right)\quad r>r_0.
\end{aligned}
\label{eq14}
\end{equation}
where $N$ is the normalization factor, $k_1$ = $\sqrt{|E^2-M_0^2|}$/$A$,
$k_2$ = $|E|$/$A$, $J_n$ and $Y_n$ are Bessel functions, and $I_n$ are
modified Bessel functions. The corresponding boundary condition at $r=r_0$
is given by
\begin{eqnarray}\nonumber
a(E+M_0)J_n(k_1r_0) = E(bJ_n(k_2r_0)+cY_n(k_2r_0)),\\
ak_1J_{n-1}(k_1r_0) = k_2(bJ_{n-1}(k_2r_0)+cY_{n-1}(k_2r_0)).
\label{eq15}
\end{eqnarray}
for $|E|$ $\geq$ $|M_0|$ and
\begin{eqnarray}\nonumber
a(E+M_0)I_n(k_1r_0) = E(bJ_n(k_2r_0)+cY_n(k_2r_0)),\\
ak_1I_{n-1}(k_1r_0) = k_2(bJ_{n-1}(k_2r_0)+cY_{n-1}(k_2r_0)).
\label{eq16}
\end{eqnarray}
for $|E|$ $<$ $|M_0|$.

For the large $r$, we generally take a cut-off, labeled by $R$, to prevent
the divergence. The asymptotic form of Bessel functions is given by
\begin{eqnarray}\nonumber
J_n(kr) \sim \sqrt{\frac{2}{\pi k r}}\cos(kr-\frac{n\pi}{2}-\frac{\pi}{4}),\\
Y_n(r) \sim \sqrt{\frac{2}{\pi k r}}\sin(kr-\frac{n\pi}{2}-\frac{\pi}{4}),
\label{eq18}
\end{eqnarray}
for a large $r$. The level spacing is given by $\Delta E$ = $A\pi/R$. As $R$
goes to infinity, the energy spectrum becomes continuous. The normalization
factor $N$ is determined by the large $r$ behavior. We can consider the
integral of $|\phi_n|^2$ in the region $[R_0,R]$, where $R\rightarrow\infty$.
We choose the parameter $R_0$ so that the Bessel functions can be well
described by their asymptotic forms when $r>R_0$ ($R_0>r_0$). As a result,
the integral of $\phi_n|^2$ in the range $[R_0,R]$ is given by
\begin{eqnarray}\nonumber
&&4\pi\int_{R_0}^R|\phi_n|^2\\
&&= \frac{4\pi}{N^2}\int_{R_0}^{R}rdrE^2[(bJ_n(k_2r)+cY_n(k_2r))^2+\nonumber\\
&&(bJ_{n-1}(k_2r)+cY_{n-1}(k_2r))^2]\nonumber\\
&&= \frac{4\pi}{N^2}\int_{R_0}^{R}rdr\frac{2}{\pi k_2r}E^2(b^2+c^2)\nonumber\\
&&= \frac{8E^2(R-R_0)}{k_2N^2}(b^2+c^2).
\label{eq19}
\end{eqnarray}
As $R$ approaches infinity, this integral will diverge linearly, thus being
dominant over the integral of $|\phi_n|^2$ in other range. Thus, the
normalization factor is determined by the integral in the range $[R_0,R]$ and
can be chosen as $\frac{1}{N}=\sqrt{\frac{1}{8|E|AR(b^2+c^2)}}$ for a
sufficient large $R$ in the numerical calculation.
With the obtained wave function, the LDOS is given by
\begin{equation}\nonumber
\begin{aligned}
&\rho(r,E) = \frac{R}{A\pi}\sum_{n}|\Psi_n(r)|^2\\
&= \frac{R}{2A\pi^2}\sum_{n}[\phi_{1,n}^2+\phi_{2,n}^2]\\
&= \frac{1}{16\pi^2 A^2|E|(b^2+c^2)}\sum_na^2[(M_0+E)^2J_n^2(k_1r)+\ldots\\
&\qquad|E^2-M_0^2|J_{n-1}^2(k_1r)]\qquad\qquad\qquad r<r_0,\\
&\ \textrm{or}\\
&= \frac{1}{16\pi^2 A^2|E|(b^2+c^2)}\sum_nE^2[bJ_n(k_2r)+cY_n(k_2r))^2+\ldots\\
&\qquad(bJ_{n-1}(k_2r)+cY_{n-1}(k_2r))^2]\qquad\quad r>r_0.
\end{aligned}
\label{eq20}
\end{equation}

To calculated the LDOS around a magnetic impurity and at $E$, there are three
free parameters in the equation of $\rho(r,E)$, i.e. $M_0$, $A$, and $r_0$.
For SmB$_6$, the Dirac cone is estimated to be around $-$5 meV with a Fermi
velocity of 1000 m/s to 10000 m/s, depending on the location of the Dirac
cone (at $\overline{X}$-point or $\overline{\Gamma}$-point). Here we focus
on the dominant peak at $-$6.5 mV, so $E$ $\approx$ $-$1.5 meV. Our STM
measurements indicate that $r_0$ $\approx$ 2.2 nm for Gd. Consequently, we
are left only with $M_0$ and $A$ as free parameters. As shown in Fig.\ 2I
of the main text, this analytical solution can reproduce our experimental
data very well, using $A$ = 3000 m/s and $M_0$ = 1.35 meV.

As mentioned in the main text, the proposed model was developed for magnetic
impurities and as such, is not necessarily expected to be applicable to
non-magnetic impurities as well. However, in order to obtain estimates for
$\ell_{sup}$ and $h_{sup}$ in a comparable fashion, we also used this model to
describe the LDOS around impurities in the pure and Y-substituted SmB$_6$.
Taking the same $E$ and $A$ values as for the Gd-case, we obtained $M_0 =$
0.7 meV and 1.35 meV for pure and Y-substituted SmB$_6$, respectively, while
$r_0$ has the same value of 1.5 nm in both cases. Also, because of the
uncertain applicability of the model in case of non-magnetic impurities
we kept a distinction between $r_0$ (the exchange interaction range in the
model) and $\ell_{sup}$ (the experimentally observed suppression of the
LDOS).

\end{document}